# Monopoles in High Temperature Phase of SU(2) QCD

Shinji Ejiri [a]

[a] Department of Physics, Kanazawa University, Kanazawa 920-11, Japan

We investigated a behavior of monopole currents in the high temperature phase of abelian projected finite temperature SU(2) QCD in maximally abelian gauge. Wrapped monopole currents which are closed by periodic boundary play an important role for the spatial string tension. And the wrapped monopole current density seems to be non-vanishing in the continuum limit. These results may be related to Polyakov's analysis of the confinement mechanism using monopole gas in 3-dimensional SU(2) gauge theory with Higgs fields.

## 1. Introduction

A characteristic feature of finite temperature QCD is the deconfinement phase transition and the string tension which is the most important quantity for quark confinement vanishes at the critical temperature. However, there are some non-perturbative quantities even in the high temperature ( deconfinement ) phase of QCD. The spatial string tension is known to be such a non-perturbative quantity. This quantity is the string tension extracted from space-like Wilson loops. The spatial string tension is non-vanishing even in the high temperature phase.

This property in the high temperature phase may be understood by dimensional reduction [1]. 4-dimensional QCD at high temperature limit can be regarded as an effective 3-dimensional QCD with Higgs fields. The effective theory has confinement features. The string tension in this effective theory may be the spatial string tension in 4-dimensional QCD. The relation between the spatial string tension and that of the effective theory is confirmed using Monte-Carlo simulations [2, 3]. The spatial string tension show the scaling behavior obtained from the effective theory

$$\sqrt{\sigma_s} \propto g^2(T)T, \qquad (1)$$

where $\sigma_s$ is the spatial string tension. $g(T)$ is 4-dimensional coupling constant.

On the other hand, the study of topological quantities ( monopole, instanton, ...) is important. Many people believe that the dual Meissner effect due to condensation of color magnetic monopoles is the color confinement mechanism in QCD [4, 5]. Polyakov showed that the confinement features of 3-dimensional SU(2) gauge theory with Higgs fields is explained by monopole gas ( 3-dimensional instanton ) analytically [6].

Last year, we studied the contribution of the monopole to the string tension in abelian projected SU(2) QCD in maximally abelian gauge, and found the following results [7].

1. Both the physical and the spatial string tension can almost be explained by monopoles alone.

2. A long monopole loop is important for the physical string tension.

3. The physical string tension and the long monopole loop disappear at the same temperature ( $T_c$ ).

4. In the high temperature phase, the total monopole number is small.

These results suggest that the important monopole loops are different in the cases of the spatial string tension and the physical one, and the spatial string tension is produced by small number of monopoles.

The aim of this report is to find what kind of monopole produces the spatial string tension in the high temperature phase of 4-dimensional QCD, and to study the relation to the Polyakov's monopole gas in the continuum theory.

## 2. Wrapped monopole loop and spatial string tension

In 3-dimensional SU(2) gauge theory with Higgs fields, monopole gas produces the string tension. When we consider dimensional reduction in 4-dimensional SU(2) gauge theory, static monopole loops correspond to monopole gas in 3-dimensional effective theory. The static monopole loops are closed by periodic boundary in the time direction. We call these loops wrapped monopole loops.

Since the temperature is the inverse of the lattice size of the time direction, the monopole currents may become easier to wrap by the periodic boundary as the temperature becomes higher. It may be related to the scaling behavior of the spatial string tension, which becomes larger as the temperature rises.

In order to know if the cause of the spatial string tension at high temperature is the monopole gas in the 3-dimensional theory, we checked two points. (1) Do only the wrapped monopole loops produce the spatial string tension? (2) How is the temperature dependence of the wrapped monopole current density? Does this value remain non-vanishing in the continuum limit?

An abelian theory is extracted by abelian projection [8, 9]. The partial gauge fixing is done in the maximally abelian gauge. An abelian link field $u(s,\mu)$ is extracted from the gauge fixed SU(2) link variable $U(s,\mu)$ as follows:

$$U(s,\mu) = c(s,\mu)u(s,\mu), \qquad (2)$$

$$u(s,\mu) = \begin{pmatrix} e^{i\theta_\mu(s)} & 0 \\ 0 & e^{-i\theta_\mu(s)} \end{pmatrix}, \qquad (3)$$

where $\theta_\mu(s)$ is the abelian gauge field. The monopole current $k_\mu(s)$ is defined as

$$\begin{aligned} k_\mu(s) &= (1/4\pi)\epsilon_{\mu\alpha\beta\gamma}\partial_\alpha\bar{\Theta}_{\beta\gamma}(s) \qquad (4) \\ \Theta_{\mu\nu}(s) &= \partial_\mu\theta_\nu(s) - \partial_\nu\theta_\mu(s) \\ \Theta_{\mu\nu}(s) &= \bar{\Theta}_{\mu\nu}(s) + 2\pi n_{\mu\nu}(s) \\ &\quad (-\pi < \bar{\Theta}_{\mu\nu} \leq \pi, \quad n_{\mu\nu} : \text{integer}) \end{aligned}$$

following Degrand-Toussaint [10].

We defined wrapping number of every cluster of connected monopole currents as follows:

$$(\text{wrapping number}) = \frac{1}{N_t}\sum_{\{\text{cluster}\}} k_4(s), \qquad (5)$$

where $\sum_{\{\text{cluster}\}}$ means summing up in a cluster, $N_t$ is the lattice size of the time direction. If a monopole current belongs to a cluster which has non-zero wrapping number, we regarded it as the wrapped monopole current.

## 3. Wrapped monopole contribution to the spatial string tension

We studied the wrapped monopole contribution to the spatial string tension in the high temperature phase. We considered the monopole contribution to the Wilson loop as discussed in [11, 12]. First, we extracted abelian component by performing abelian projection in the maximally abelian gauge. In this gauge, the spatial string tension can be reproduced by residual abelian link variables. Next, we decomposed the abelian Wilson loop $W$, which is the Wilson loop composed of abelian link variables, into two parts $W_1$ ( photon part ), $W_2$ ( monopole part ) as follows:

$$W = \exp\{i\sum \theta_\mu(s)J_\mu(s)\} \qquad (6)$$
$$W = W_1 \cdot W_2 \qquad (7)$$
$$W_1 = \exp\{-i\sum \partial'_\mu \bar{\Theta}_{\mu\nu}(s)D(s-s')J_\nu(s')\}$$
$$W_2 = \exp\{2\pi i\sum k_\beta(s)D(s-s')\frac{1}{2}\epsilon_{\alpha\beta\rho\sigma}\partial_\alpha M_{\rho\sigma}(s')\},$$

where $D(s-s')$ is the lattice Coulomb propagator. $J_\mu(s)$ is an external current corresponding to the Wilson loop and $M_{\mu\nu}(s)$ is an antisymmetric variable taking $\pm 1$ on a surface with the Wilson loop boundary as $J_\nu(s) = \partial'_\mu M_{\mu\nu}(s)$. Notice that we can see that a time-like monopole ($k_4$) such as the static monopole does not affect the physical string tension and contributes only to the spatial one as seen from this equation.

The monopole contribution in deconfinement phase is a little bit lower than the full one, but it almost reproduce the behavior of the full one in the maximally abelian gauge [7].

Here, we calculated the wrapped monopole contribution and the non-wrapped monopole contribution to the spatial string tension separately in the high temperature phase.

We performed Monte-Carlo simulations on $24^3 \times N_t$ lattices, $N_t = \{2,4,6,8\}$, at $\beta =$

$\{2.30, 2.51, 2.74\}$ which are the critical $\beta$ for $N_t = \{4, 8, 16\}$ respectively. Measurements were done every 50 sweeps after a thermalization of 2000 sweeps. We took 50 configurations for measurements. The data are plotted in Fig. 1.

These data show that the spatial string tension from the wrapped monopole is almost the same as that from total monopole loops and that the non-wrapped loops do not contribute to the spatial string tension.

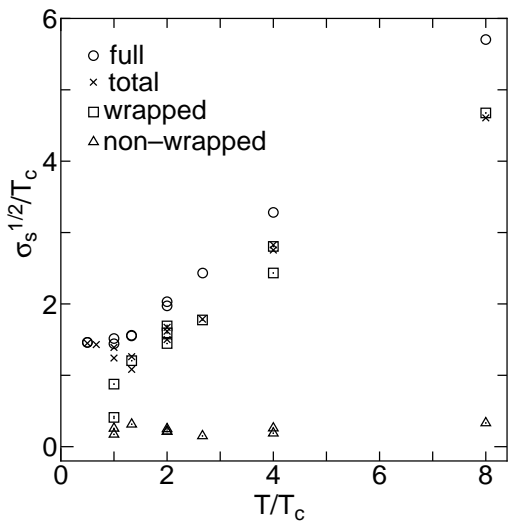

Figure 1. The full spatial string tension (circle), the total monopole contribution (cross), the wrapped monopole contribution (square) and the non-wrapped monopole contribution (triangle). The full one is cited from [2]

## 4. Scaling behavior of wrapped monopole density

If the wrapped monopole produces the spatial string tension in the continuum limit, the wrapped monopole density must remain non-vanishing in the continuum limit. The monopole density $\rho(T)$ on a lattice of size $N_s^3 \times N_t$ defined as follows:

$$\rho(T) = \frac{\sum |k_\mu(s)| a}{(N_s a)^3 (N_t a)}. \quad (8)$$

Here $a$ is the lattice spacing. Considering the relation between the temperature and the lattice size: $N_t a = \frac{1}{T}$, we can rewrite the monopole density as follows:

$$\rho(T) = \frac{\sum |k_\mu(s)| N_{tc}}{N_s^3 N_t} T_c^3, \quad (9)$$

where $T_c$ is the critical temperature. $N_{tc}$ is the critical lattice size at every $\beta$.

We measured the temperature dependence and the $\beta$ dependence of the total monopole density and the wrapped monopole density varying both $\beta$ and $N_t$ on $24^3 \times N_t$ lattices. ($\beta = \{2.30, 2.51, 2.74\}$, $N_t = \{2, 4, 6, 8, 12\}$)

The data of total and wrapped monopole densities are shown in Fig. 2 and in Fig. 3.

The total monopole density dose not show good scaling behavior. It depends not only on $T$ but also on $\beta$. On the other hand, the wrapped monopole density in the high temperature phase is independent of $\beta$, and seems to remain in the continuum limit. The wrapped monopole density seems to be proportional to $T^3$ at high temperature.

## 5. Conclusions and Discussion

We found the following results by the numerical studies of Monte-Carlo simulations. The spatial string tension can almost be reproduced by the wrapped monopole loops closed by the periodic boundary in the time direction in the high temperature phase. The wrapped monopole density is independent of $\beta$ and remains in the continuum limit. Moreover the scaling behavior of this value is similar to that of the spatial string tension. These results suggest that the spatial string tension at high temperature is produced by the monopole gas in the effective 3-dimensional theory.

In Polyakov's analytical calculation, it is essential that the Higgs field has non-zero expectation value. On the other hand, the study of the scaling behavior of the spatial string tension [2, 3] suggested that the Higgs-sector in dimensionally reduced QCD does not contribute significantly to the spatial string tension. We expect that the monopole gas can be discussed without the expectation value of the Higgs fields as we are discussing the monopole currents in 4-dimensional QCD without Higgs fields. Although we do not know if the Higgs field in the effective 3-dimensional theory has non-vanishing vacuum expectation value, these results suggest that the spatial string tension has a close relation to the monopole gas.

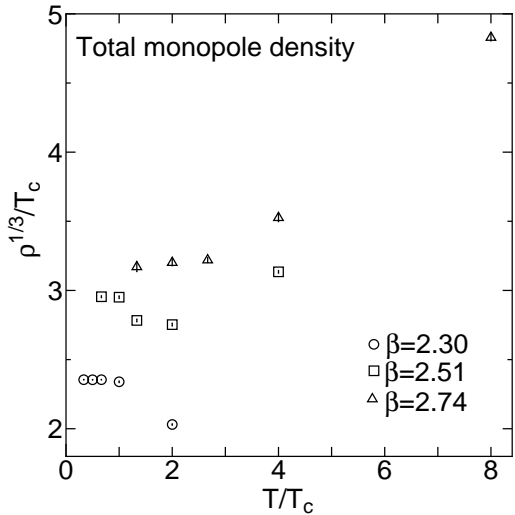

Figure 2. The temperature dependence of the total monopole density at $\beta = 2.30$ (circle), $\beta = 2.51$ (square) and $\beta = 2.74$ (triangle) on $N_t = 2, 4, 6, 8$ and 12 lattices respectively.

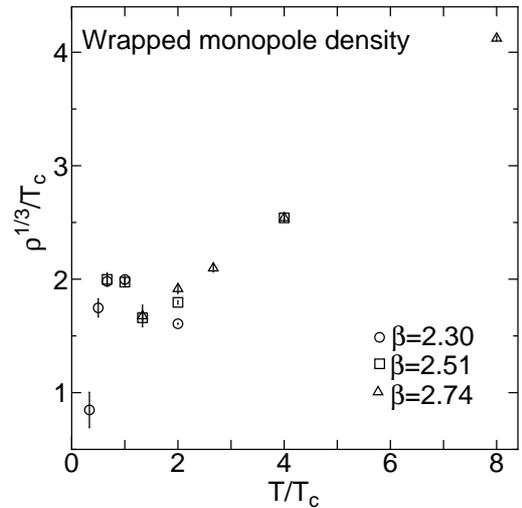

Figure 3. The temperature dependence of the wrapped monopole density at $\beta = 2.30$ (circle), $\beta = 2.51$ (square) and $\beta = 2.74$ (triangle) on $N_t = 2, 4, 6, 8$ and 12 lattices respectively.

**Acknowledgments**

The author thanks T.Suzuki, Y.Matsubara and S.Kitahara for fruitful discussions and comments. Prof. T.Suzuki suggested that the spatial string tension at high temperature has a relation to the Polyakov's monopole gas. The calculations were performed on Fujitsu VPP500 at the institute of Physical and Chemical Research (RIKEN) and National Laboratory for High Energy Physics (KEK).